\documentclass[twocolumn, graphics, floatfix, a4paper, aps, prb,
superscriptaddress, longbibliography, showpacs,citeautoscript]{revtex4-1}

\usepackage{graphicx}
\usepackage{amsmath}
\usepackage{amssymb}
\usepackage{color}
\usepackage{ulem}
\usepackage[dvipsnames]{xcolor}
\usepackage[colorlinks=true,urlcolor=blue,citecolor=blue,linkcolor = blue]{hyperref}
\usepackage{breakurl}

\hypersetup{breaklinks=true}

\renewcommand{\vec}[1]{\boldsymbol{#1}}

\newcommand{\crea}[1]{#1^{\dag}}
\newcommand{\anni}[1]{#1^{\vphantom{\dag}}}

\newcommand{\bra}[1]{\langle #1|}
\newcommand{\ket}[1]{|#1\rangle}
\newcommand{\braket}[2]{\langle #1|#2\rangle}

\newcommand{\bea}{\begin{equation} \begin{aligned}}
\newcommand{\eea}{\end{aligned} \end{equation} }

\newcommand{\eq}[1]{\begin{equation}#1\end{equation}}
\newcommand{\eqa}[1]{\begin{align}#1\end{align}}

\begin{document}

\title{Conductivity in flat bands from the Kubo-Greenwood formula}
\author{Kukka-Emilia Huhtinen}
\email{kukka-emilia.huhtinen@aalto.fi}
\affiliation{Department of Applied Physics, Aalto University School of Science,
FI-00076 Aalto, Finland}
\author{P\"aivi T\"orm\"a}
\email{paivi.torma@aalto.fi}
\affiliation{Department of Applied Physics, Aalto University School of Science,
FI-00076 Aalto, Finland}

\date{\today}

\begin{abstract}
Conductivity in a multiband system can be divided into intra- and interband contributions, and the latter further into symmetric and antisymmetric parts. In a flat band, intraband conductivity vanishes and the antisymmetric interband contribution, proportional to the Berry curvature, corresponds to the anomalous Hall effect. We investigate whether the symmetric interband conductivity, related to the quantum metric, can be finite in the zero frequency and flat band limit. Starting from the Kubo-Greenwood formula with a finite scattering rate $\eta$, we show that the DC conductivity is zero in a flat band when taking the clean limit ($\eta \rightarrow 0$). If commonly used approximations involving derivatives of the Fermi distribution are used, finite conductivity appears at zero temperature $T=0$, we show however that this is an artifact due to the lack of Fermi surfaces in a (partially) flat band. We then analyze the DC conductivity using the Kubo-Streda formula, and note similar problems at $T=0$. The predictions of the Kubo-Greenwood formula (without the approximation) and the Kubo-Streda formula differ significantly at low temperatures. We illustrate the results within the Su-Schrieffer-Heeger model where one expects vanishing DC conductivity in the dimerized limit as the unit cells are disconnected. We discuss the implications of our results to previous work which has proposed the possibility of finite flat-band DC conductivity proportional to the quantum metric. Our results also highlight that care should be taken when applying established transport and linear response approaches in the flat band context, since many of them utilize the existence of a Fermi surface and assume scattering to be weak compared to kinetic energy.   
\end{abstract}

\maketitle

\section{Introduction}

Quantum geometry is key to understanding multiband systems. For
instance, quantum geometry has been related to superconductivity in
flat
bands~\cite{Peotta2015,Liang2017,Julku2016,Rossi2021,Torma2018,Julku2020,Hu2019,Xie2020,Iskin2021,Huhtinen2022,Iskin2022},
orbital magnetic 
susceptibility~\cite{Piechon2016,Gao2015}, 
light-matter interactions~\cite{Holder2020,Topp2021}, the 
intrinsic anomalous Hall effect~\cite{Thouless1982,Niu1985,Kohmoto1985,Onoda2002,Jungwirth2002}, and other physical
phenomena~\cite{Rhim2020,Abouelkomsan2022,Gao2019,Ahn2021,Iskin2020,Julku2021,Gianfrate:2019}. Quantum geometric quantities determine the phase and amplitude distances between quantum states, and are represented by the quantum geometric tensor~\cite{Provost1980} whose imaginary (antisymmetric) part is the Berry curvature and real (symmetric) part the quantum metric (Fubini-Study metric). The quantum metric and Berry curvature are
particularly central in the properties of flat bands systems. Flat bands are interesting platforms for strongly correlated quantum phenomena, and have
attracted increased interest due to their relevance in moiré
materials~\cite{Cao2018,Cao2018b,Yankowitz2019,Park2021,Shen2020,Cao2020,lu:2019,Tian2021,Balents2020,Andrei2021,Torma2021}.

Non-interacting particles on flat bands are localized and have a
diverging effective mass. However, recent results have predicted a
nonzero DC conductivity weakly sensitive to the inelastic scattering
rate~\cite{Bouzerar2020,Bouzerar2021}, also found in disorder-induced quasilocalized
zero-energy modes in graphene~\cite{Ferreira2015}. This result is
sensitive to the used approach, and for instance wave-packet
propagation methods predict a vanishing conductivity of the
zero-energy modes~\cite{Fan2014,Cresti2013}. A nonzero or even
diverging DC 
conductivity has also been  
predicted in disordered non-isolated flat
bands~\cite{Vigh2013,Wang2020}. In perfectly
flat bands, the zero-temperature DC conductivity has been derived to be
proportional to the quantum
metric~\cite{Mitscherling2022,Mera2022,Bouzerar2022}.

The conductivity is often computed using the Kubo-Greenwood formula,
which is the non-interacting version of the exact Kubo formula. The
Kubo-Greenwood formula can be sensitive to the order the relevant limits (zero temperature, zero scattering, zero frequency) are taken~\cite{Ziegler2007} or to approximations made for instance in the
delta 
functions~\cite{Calderin2017}. Here, we show that when applied to flat bands, the
Kubo-Streda formula can give drastically different results than
the Kubo-Greenwood formula obtained directly via an independent
particle approximation of the Kubo formula. In particular, if the zero
temperature limit is taken before taking the scattering rate to zero,
it can yield a conductivity proportional to the quantum metric in the
clean limit, which does not appear when applying the Kubo-Greenwood
formula which predicts vanishing DC conductivity.

In Section~\ref{sec.kg}, we first derive the conductivity within the Kubo-Greenwood formula dividing it into intra- and interband contributions and their symmetric and antisymmetric parts. We show how in a flat band, where the intraband conductivity is zero and the antisymmetric interband contributions give the anomalous Hall effect, the symmetric intraband contribution vanishes at small temperatures in the clean limit. This means no DC conductivity. In Section~\ref{sec.comparison}, we then compute the conductivity using the Kubo-Streda formula, and show, using also the sawtooth ladder and the dimerized SSH chain as examples, that the results differ dramatically from the Kubo-Greenwood formula at low temperatures. We explain the origin of the discrepancy. In Section~\ref{sec.conclusion}, we summarize and discuss our results. 

\section{Kubo-Greenwood formula} \label{sec.kg}

\begin{figure*}
\centering
\includegraphics[width=\textwidth]{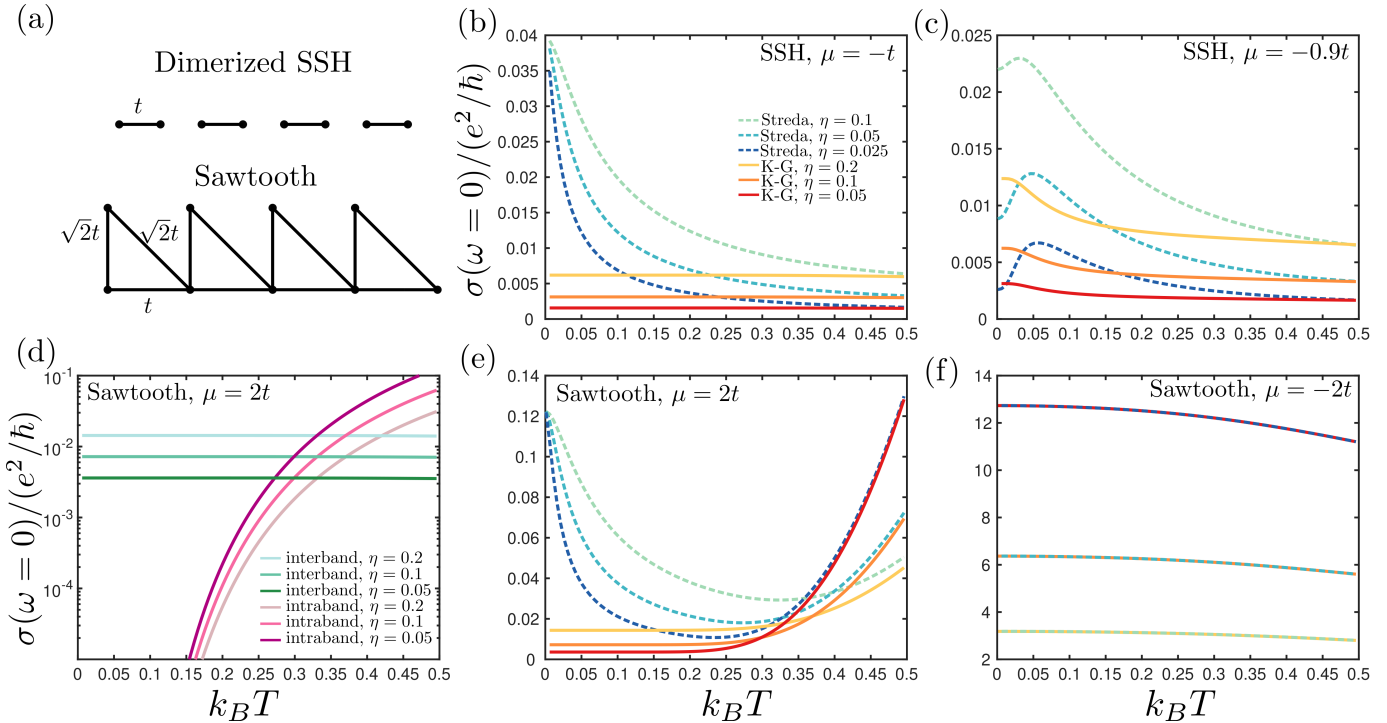}
\caption{(a) Sketch of the dimerized SSH model and the sawtooth ladder studied here. (b-c) DC conductivity in the SSH model at $\mu=-t$, when the chemical potential is in the lowest flat band (b), and at $\mu=-0.9t$ (c). The conductivity obtained from the Kubo-Greenwood formula consists of solely interband contributions, and vanishes at all temperatures in the clean limit $\eta\to 0^+$. The DC conductivity obtained from the Kubo-Streda formula remains pinned at $1/(8\pi)$ at $T=0$, but vanishes for nonzero temperatures. When the chemical potential is tuned away from the flat band, the limit $\eta\to 0^+$ from the Kubo-Greenwood and Kubo-Streda formulae is the same, although the behavior at nonzero $\eta$ is drastically different. (d) Interband and intraband contributions obtained from the Kubo-Greenwood formula in the sawtooth ladder when the chemical potential is in the flat band. The intraband contribution from the dispersive band diverges as $\eta\to 0^+$, whereas the interband contribution vanishes. (e-f) DC conductivity from the Kubo-Greenwood and Kubo-Streda formula when the chemical potential is (e) in the flat band and (f) in the middle of the dispersive band. In a dispersive band, both methods give the same results, whereas in a flat band, they give drastically different results. In the flat band, the conductivity from the Kubo-Streda formula remains pinned to $2/(3\sqrt{3}\pi)$ at $T=0$.}
\label{fig}
\end{figure*}

We consider fermionic multiband models described by the tight-binding
Hamiltonian 
$H=\sum_{i\alpha,j\beta} t_{i\alpha,j\beta}
\crea{c_{i\alpha}}\anni{c_{j\beta}} - \mu \sum_{i\alpha}
n_{i\alpha}$, where $t_{i\alpha,j\beta}$ is the hopping amplitude from
site $j\beta$ to $i\alpha$. The unit cells are labeled with $i,j$, while
$\alpha,\beta$ indicate the orbitals within a unit cell. By taking the
Fourier transformation $c_{i\alpha}
= (1/\sqrt{N_c})\sum_{\vec{k}}
c_{\vec{k}\alpha}e^{i\vec{k}\cdot(\vec{R}_i+\vec{\delta}_{\alpha})}$
the Hamiltonian becomes $H=\sum_{\vec{k}} H_{\vec{k}}$, where 
\eq{
  H_{\vec{k}} = \sum_{i}\sum_{\alpha\beta} t_{i\alpha,0\beta}
  e^{-i\vec{k}\cdot(\vec{R}_i + \vec{\delta}_{\alpha}-\vec{\delta}_{\beta})}.
}
Here, $\vec{R}_i$ is the position of the $i$:th unit cell, and
$\vec{\delta}_{\alpha} = \vec{r}_{i\alpha}-\vec{R}_i$, with
$\vec{r}_{i\alpha}$ the position of site $i\alpha$. 
The eigenvalues and eigenvectors give the band dispersion relations
$\epsilon_n(\vec{k})$ and the periodic parts of the Bloch functions
$\ket{n_{\vec{k}}}$, respectively.

Note that in a
multiband lattice taking the orbital positions $\vec{\delta}_{\alpha}$ into account in the
Fourier transformation is essential in order to obtain the correct
conductivity~\cite{Tomczak2009,Nourafkan2018,Mitscherling2020}. This is in contrast to the superfluid weight,
which is independent of the particular choice of
$\vec{\delta}_{\alpha}$ provided it is computed accurately~\cite{Huhtinen2022}. In
other words, the conductivity is generally geometry-dependent, using
the terminology introduced in Ref.~\onlinecite{Simon2020}.  

The conductivity tensor $\sigma_{ij}(\omega)\equiv
\sigma_{ij}(\omega,\vec{q}=\vec{0})$ in a 
non-interacting system is given by the Kubo-Greenwood formula~\cite{Greenwood1958} 
\eqa{
  \sigma_{\mu\nu}(\omega) &= \frac{e^2}{i\hbar V}\sum_{\vec{k}}\sum_{mn}
  \frac{n_F(\epsilon_n(\vec{k}))-n_F(\epsilon_m(\vec{k}))}{\epsilon_n(\vec{k})-\epsilon_m(\vec{k})}
  \times \nonumber\\
  &\times \frac{[j_{\mu}(\vec{k})]_{nm}[j_{\nu}(\vec{k})]_{mn}}{\epsilon_n(\vec{k})-\epsilon_m(\vec{k})+\hbar
    \omega+ i\eta}, \label{eq.kg_true}
}
which is obtained from the Kubo formula~\cite{Kubo1957} by performing an
independent electron approximation. The prefactor involving the
Fermi-Dirac distribution $n_F(\epsilon)=1/(e^{\beta \epsilon}+1)$,
with $\beta=1/(k_BT)$, should be understood as
$\partial_{\epsilon}n_F(\epsilon)|_{\epsilon=\epsilon_n}$ when
$\epsilon_n(\vec{k})=\epsilon_m(\vec{k})$~\cite{Calderin2017}. The infinitesimal imaginary
shift $\eta$ added to the frequency acts as a
small inelastic scattering rate or relaxation rate. The
current operators $j_{\mu}$ are obtained from the momentum derivatives
of $H_{\vec{k}}$, 
$j_{\mu}(\vec{k}) = \partial_{k_{\mu}}H_{\vec{k}}$, and
\eq{
  [j_{\mu}(\vec{k})]_{mn} =
\partial_{k_{\mu}}\epsilon_m(\vec{k})\delta_{mn} +
(\epsilon_m(\vec{k})-\epsilon_n(\vec{k}))
\braket{\partial_{k_{\mu}}m_{\vec{k}}}{n_{\vec{k}}}. \label{eq.current_ops}
}

Another widely
used form of the Kubo-Greenwood formula for the diagonal components of
$\sigma_{\mu\nu}$ is
\eqa{
  \sigma_{\mu\mu} &= -\frac{e^2}{\hbar \pi V} \sum_{\vec{k}}\sum_{mn}
  \int_{-\infty}^{\infty} d\epsilon \: \frac{\partial
    n_F(\epsilon)}{\partial \epsilon}\times \nonumber\\
    &\times{\rm Tr} [{\rm
      Im}[G_{\vec{k}}(\epsilon+i\eta)] j_{\mu}(\vec{k}) {\rm
      Im}[G_{\vec{k}}(\epsilon+i\eta)]
    j_{\mu}(\vec{k})], \label{eq.kg_wrong} 
}
where $G_{\vec{k}}(E) = (E-H_{\vec{k}})^{-1}$ is the Green's function. This is simply the
diagonal components of the more general Bastin~\cite{Bastin1971} and Streda~\cite{Streda1982} formulae. 
In this work, we show that when applied to flat bands,
Eqs.~\eqref{eq.kg_wrong} 
and~\eqref{eq.kg_true} give drastically different results at $T=0$. In
order to avoid confusion between the two forms, we will refer to
Eq.~\eqref{eq.kg_true} as the Kubo-Greenwood formula, and to
Eq.~\eqref{eq.kg_wrong} as the Kubo-Streda formula. 

Let us first derive ${\rm Re}[\sigma_{\mu\nu}]$ from
Eq.~\eqref{eq.kg_true}. 
The real part of the conductivity can be decomposed in several ways,
for 
instance into so-called Fermi surface and Fermi sea
contributions~\cite{Streda1982,Crepieux2001,Bonbien2020}. In our case, we choose to split
$\sigma_{\mu\nu}$ to intraband, symmetric interband and antisymmetric
interband contributions, similarly to the decomposition used in
Ref.~\onlinecite{Mitscherling2020}. The advantage of this split when considering flat bands
is apparent: the intraband contribution from a perfectly flat band
vanishes exactly, and only the interband part remains. The
antisymmetric part of the latter is related to the intrinsic anomalous
Hall effect.

In the thermodynamic limit, the intraband contribution to the real
part of the conductivity 
$\sigma^{\rm intra}_{\mu\nu}$ obtained from Eq.~\eqref{eq.kg_true} is
\eqa{
  {\rm Re}\: \sigma^{\rm intra}_{\mu\nu}(\omega) &= -\frac{e^2}{\hbar}
  \sum_{n}\int_{\rm B.z.} \frac{d^D\vec{k}}{(2\pi)^D}\:
  \frac{\partial 
    n_F(E)}{\partial 
    E}\bigg|_{E=\epsilon_n(\vec{k})}\times \nonumber\\
  &\times[j_{\mu}(\vec{k})]_{nn}[j_{\nu}(\vec{k})]_{nn} \frac{\eta}{(\hbar
    \omega)^2+\eta^2} \\
  &= -\frac{e^2}{\hbar} \sum_{n}\int_{\rm B.z.} \frac{d^D \vec{k}}{(2\pi)^D}
  \frac{\partial 
    n_F(E)}{\partial 
    E}\bigg|_{E=\epsilon_n(\vec{k})} \times \nonumber \\
  &\times\partial_{k_{\mu}}\epsilon_n(\vec{k})
  \partial_{k_{\nu}}\epsilon_n(\vec{k})
  \frac{\eta}{(\hbar\omega)^2+\eta^2}.
}
We have replaced the momentum summation by an integral over the first
Brillouin zone, $(1/V)\sum_{\vec{k}}\to (1/2\pi)^D\int_{\rm
  B.z.}d^D\vec{k}$, where $D$ is the dimension of the system.
This contribution to the conductivity is the only component present in a single-band model, and gives
the same result as the semiclassical Boltzmann theory of transport when taking $\tau=1/\eta$ as a momentum-independent relaxation time. The
intraband contribution is clearly zero in a perfectly dispersionless
band. 

The total interband contribution is
\begin{widetext}
\eq{
  \sigma_{\mu\nu}^{\rm inter} = -i\frac{e^2}{\hbar}\sum_{m\neq n} \int_{\rm B.z.}
  \frac{d^D\vec{k}}{(2\pi)^D}
  \frac{n_F(\epsilon_n(\vec{k}))}{\epsilon_n(\vec{k})-
    \epsilon_m(\vec{k})} \left( 
  \frac{[j_{\mu}(\vec{k})]_{nm}[j_{\nu}(\vec{k})]_{mn}}{\epsilon_n(\vec{k})
    - \epsilon_m(\vec{k}) +\hbar \omega + i\eta} +
  \frac{[j_{\nu}(\vec{k})]_{nm}[j_{\mu}(\vec{k})]_{mn}}{\epsilon_m(\vec{k})
    - \epsilon_n(\vec{k}) +\hbar \omega + i\eta}
\right).
}
Using Eq.~\eqref{eq.current_ops}, we can express the symmetric and
antisymmetric parts of the real part of $\sigma_{\mu\nu}^{\rm inter}$
as
\eqa{
  \sigma^{s}_{\mu\nu}(\omega)&=-\frac{e^2}{\hbar} \sum_{n\neq m}\int_{\rm B.z}
  \frac{d^D\vec{k}}{(2\pi)^D} n_F(\epsilon_n(\vec{k})) 
  {\rm
    Re}[\braket{\partial_{k_{\mu}}n_{\vec{k}}}{m_{\vec{k}}}\braket{m}{\partial_{k_{\nu}}n_{\vec{k}}}] 
  \times\nonumber \\
  &\times\left(
    \frac{\eta(\epsilon_n(\vec{k})-\epsilon_m(\vec{k}))}{(\epsilon_n(\vec{k})-\epsilon_m(\vec{k})+\hbar\omega)^2+\eta^2}
    +
    \frac{\eta(\epsilon_n(\vec{k})-\epsilon_m(\vec{k}))}{(\epsilon_n(\vec{k})-\epsilon_m(\vec{k})-\hbar\omega)^2+\eta^2}
  \right),\\
  \sigma^a_{\mu\nu}(\omega)&=\frac{e^2}{\hbar} \sum_{\vec{k}}\sum_{n\neq m}\int_{\rm
    B.z.} \frac{d^D\vec{k}}{(2\pi)^D}
  n_F(\epsilon_n(\vec{k})) 
  {\rm
    Im}[\braket{\partial_{k_{\mu}}n_{\vec{k}}}{m}\braket{m_{\vec{k}}}{\partial_{k_{\nu}}n_{\vec{k}}}] 
  \times\nonumber \\
  &\times\left(
    \frac{(\epsilon_n(\vec{k})-\epsilon_m(\vec{k}))(\epsilon_n(\vec{k})-\epsilon_m(\vec{k})
      +
      \hbar\omega)}{(\epsilon_n(\vec{k})-\epsilon_m(\vec{k})+\hbar\omega)^2+\eta^2} 
    +
    \frac{(\epsilon_n(\vec{k})-\epsilon_m(\vec{k}))(\epsilon_n(\vec{k})-\epsilon_m(\vec{k})
      -
      \hbar\omega)}{(\epsilon_n(\vec{k})-\epsilon_m(\vec{k})-\hbar\omega)^2+\eta^2}
  \right),
}
\end{widetext}

The interband contribution is thus related to components of the
quantum geometric tensor~\cite{Resta2011}, which for the $n$:th band
is defined as
\eqa{
  \mathcal{G}^n_{\mu\nu}(\vec{k}) &=  2
  \bra{\partial_{k_{\mu}}n_{\vec{k}}}\left(1-\ket{n_{\vec{k}}}\bra{n_{\vec{k}}}\right)\ket{\partial_{k_{\nu}}n_{\vec{k}}}
  \\
  &= 2 \sum_{m:m\neq n} \braket{\partial_{k_{\mu}}n_{\vec{k}}}{m_{\vec{k}}}\braket{m_{\vec{k}}}{\partial_{k_{\nu}}n_{\vec{k}}}.
}
The real and imaginary parts of $\mathcal{G}_{\mu\nu}^n=
\mathcal{M}_{\mu\nu}^n + i\mathcal{B}_{\mu\nu}^n$ are the quantum metric
and Berry curvature, respectively. Since the quantum geometric tensor is Hermitian, the quantum metric is symmetric while the Berry curvature is antisymmetric.

In the limit $\omega\to 0$, $\eta\to 0^+$, we recover for $\sigma^a_{\mu\nu}$
the well-known relationship between the intrinsic anomalous Hall
conductivity and the Berry curvature~\cite{Thouless1982,Niu1985,Kohmoto1985,Onoda2002,Jungwirth2002,Xiao2010},
\eq{
  \sigma^a_{\mu\nu}(\omega=0) = \frac{e^2}{\hbar}\sum_n\int_{\rm B.z.}
  \frac{d^D\vec{k}}{(2\pi)^D}
  n_F(\epsilon_n(\vec{k}))\mathcal{B}_{\mu\nu}^n(\vec{k}). 
}
Notably, the antisymmetric part of the conductivity can be nonzero
even on a flat band in the clean limit $\eta\to 0^+$. In contrast to
for instance 
the superfluid weight and Drude weight, it also generally depends on
the choice of orbital positions within a unit cell~\cite{Simon2020,Huhtinen2022}. 

The symmetric interband contribution, on the other hand, is related to
components of the quantum metric, but less directly. In the limit
$\eta\to 0^+$, $\eta/(x^2+\eta^2)\to \pi\delta(x)$, and the
symmetric interband conductivity becomes
\begin{widetext}
\eqa{
  \sigma_{\mu\nu}^s(\omega) = -\frac{e^2\pi}{\hbar}\sum_{n\neq m}\int_{\rm B.z.}
  &n_F(\epsilon_n(\vec{k})) {\rm Re}\left[
    \braket{\partial_{k_{\mu}}n_{\vec{k}}}{m_{\vec{k}}}\braket{m_{\vec{k}}}{\partial_{k_{\nu}}n_{\vec{k}}} 
    \right] (\epsilon_n(\vec{k})-\epsilon_m(\vec{k}))
  \times\nonumber\\
  &\times\left(
  \delta(\epsilon_n(\vec{k})-\epsilon_m(\vec{k})+\hbar\omega)
  + \delta(\epsilon_n(\vec{k})-\epsilon_m(\vec{k})-\hbar\omega)
  \right). \label{eq.inter_kg}
}
\end{widetext}
When all bands are isolated from each other,
$|\epsilon_n-\epsilon_m|\geq E_{\rm gap,min}$ for $m\neq n$, where $E_{\rm gap,min}$
is the smallest interband gap. One can see from the delta-functions in Equation~\eqref{eq.inter_kg} that the interband conductivity then vanishes for any
frequency $\hbar\omega<E_{\rm gap,min}$ in the clean
limit. It follows that the symmetric part of the DC conductivity in an
isolated flat band is zero in the limit $\eta\to 0^+$. At nonzero
scattering rate, the DC conductivity can acquire a nonzero value
because of the spread of the Lorentzian functions centered at frequencies
$\hbar\omega\geq E_{\rm gap,min}$; this is just the finite linewidth of interband transitions resonant with an AC field. For $\eta\ll E_{\rm
  gap,min}$, this nonzero interband contribution is approximately
\eqa{
  \sigma^s_{\mu\nu}(\omega=0)&\approx -2\eta \frac{e^2}{\hbar}\sum_{n}\int_{\rm B.z.}
  \frac{d^D\vec{k}}{(2\pi)^D} n_F(\epsilon_n(\vec{k})) \times \nonumber \\
  &\times\sum_{m:m\neq
    n} 
  \frac{{\rm
    Re}[\braket{\partial_{k_{\mu}}n_{\vec{k}}}{m_{\vec{k}}}\braket{m_{\vec{k}}}{\partial_{k_{\nu}}n_{\vec{k}}}]}{\epsilon_n(\vec{k})-\epsilon_m(\vec{k}) 
  }, \label{eq.dc_kg}
}
where we have approximated $\eta
(\epsilon_n(\vec{k})-\epsilon_m(\vec{k}))/[(\epsilon_n(\vec{k}-\epsilon_m(\vec{k})))^2+\eta^2]\approx
\eta/(\epsilon_n(\vec{k})-\epsilon_m(\vec{k}))$. The linear dependence on the scattering rate is consistent with the result obtained in Refs.~\onlinecite{Mitscherling2020,Mitscherling2022} for small scattering rates in a dispersive band. Here, this holds also for perfectly flat bands. We note that an
expression which relates the DC conductivity directly to the quantum
metric can be obtained from Eq.~\eqref{eq.inter_kg} by using that
\eqa{
&\frac{n_F(\epsilon_n)-n_F(\epsilon_m)}{\epsilon_n-\epsilon_m}\delta(\epsilon_n-\epsilon_m+\hbar\omega) \nonumber \\
&= \frac{n_F(\epsilon_n+\hbar\omega)-n_F(\epsilon_n)}{\hbar\omega}
\delta(\epsilon_n-\epsilon_m+\hbar\omega)
  . \label{eq.app_fd}
}
 The prefactor on the second line should be, as mentioned earlier, understood as the derivative $\partial n_F(E)/\partial E|_{E=\epsilon_n(\vec{k})}$ when $\omega \rightarrow 0$. If the
delta functions are replaced by Lorentzian functions after this
substitution, the DC conductivity for $\eta\ll E_{\rm
  gap,min}$ becomes
\eqa{
  \sigma^s_{\mu\nu}(\omega=0) &= -\eta\frac{e^2}{\hbar}\sum_{n}\int_{\rm B.z.}
  \frac{d^D\vec{k}}{(2\pi)^D}  
  \frac{\partial n_F(E)}{\partial E}\bigg|_{E=\epsilon_n(\vec{k})}
  \times\nonumber\\
  &\times
  \sum_{m\neq n} {\rm Re}[
    \braket{\partial_{k_{\mu}}n}{m}\braket{m}{\partial_{k_{\nu}}n}].
\label{eq.sym_inter_app}
}
This form is more reminiscent of results obtained for instance by computing the conductivity in Matsubara space~\cite{Mitscherling2020,Mitscherling2022}. It should however be stressed
that here this is an approximation, as 
Eq.~\eqref{eq.app_fd} no longer holds if the delta-function is
replaced by a Lorentzian function with a finite spread. In the limit
$\omega\to 0$, the prefactor on the right-hand side of Equation~\eqref{eq.app_fd} is 
replaced by a derivative of the Fermi-Dirac distribution at
$\epsilon_n$, which approximates
$(n_F(\epsilon_n)-n_F(\epsilon_m))/(\epsilon_n-\epsilon_m)$ on the left hand side only when
$\epsilon_n$ and $\epsilon_m$ are close. If the bands are
well isolated, this does not hold. Equation~\eqref{eq.sym_inter_app}
is also problematic at $T=0$ if the Fermi energy is in a perfectly
flat band, as the derivative of the Fermi-Dirac distribution would
then become $-\delta(E_{FB}-\mu)=-\delta(0)$, where $E_{FB}$ is the energy of the flat band, at all points of the
Brillouin zone. Transforming the integral over the momentum to an integral over the energy does not help, as it would involve the density of states on the flat band, which diverges. However, at any nonzero temperature,
equations~\eqref{eq.sym_inter_app} and~\eqref{eq.dc_kg} always give
the same vanishing interband DC conductivity in the $\eta\to 0^+$
limit. 

A vanishing DC conductivity on a noninteracting isolated flat band is
unsurprising, since single particles are localized. However, it
contrasts with recent results at $T=0$, which have found a nonzero DC
conductivity proportional to the quantum metric even in the limit
$\eta\to 0^+$~\cite{Mitscherling2022,Mera2022,Bouzerar2021,Bouzerar2022}. In~\onlinecite{Mitscherling2020,Mitscherling2022}, it was shown that the interband contribution vanishes linearly with $\eta$ on dispersive bands, and this finite conductivity in the clean limit thus appears only in perfectly flat bands. In the following, we show that such
results 
can arise when applying the Kubo-Streda formula when the Fermi energy
is in a (partially) flat band. This is related to contributions from
states at exactly the Fermi energy which do not
vanish in the clean limit, present in the Streda formula but absent 
in the Kubo-Greenwood formula (Eq.~\eqref{eq.kg_true}). These
contributions only become meaningful in systems without a Fermi
surface, such as a flat band. 

\section{Flat band conductivity from the Kubo-Streda formula}\label{sec.comparison}

The Kubo-Streda formula gives the symmetric part of the DC
conductivity as~\cite{Bonbien2020,Crepieux2001,Streda1982}
\eqa{
  &\sigma_{\mu\nu}^{\rm sym} = -\frac{e^2}{\hbar\pi} 
  \int_{-\infty}^{\infty}
        {\rm d}\epsilon \frac{\partial
          n_F(\epsilon)}{\partial\epsilon} \nonumber \\
          &{\rm Tr}[{\rm
      Im}[G_{\vec{k}}(\epsilon+i\eta)] j_{\mu}(\vec{k}) {\rm
      Im}[G_{\vec{k}}(\epsilon+i\eta)]
    j_{\mu}(\vec{k})]. \label{eq.bastin}
}
This equation can be derived from the Kubo-Greenwood formula directly,
or can be obtained from the exact Kubo formula by computing the
current-current response function in Matsubara space when vertex
corrections are ignored. When applied to dispersive bands,
Equations~\eqref{eq.kg_true} and~\eqref{eq.bastin} usually give
very close results, especially at low $\eta$, provided we take
$\eta_{\rm K-G}=2\eta_{\rm Streda}$. However, when applied to flat
bands, the DC conductivities can differ drastically especially at low
temperatures.

To illustrate this, we consider two one-dimensional flat band systems:
the sawtooth ladder and the dimerized limit of the
Su-Schrieffer-Heeger (SSH) model (see Fig.~\ref{fig}a). The
sawtooth ladder features a perfectly flat band at energy $E=2t$, isolated
from a dispersive band $\epsilon(\vec{k})=-(2+2\cos(k))t$. The dimerized SSH model has two exactly flat
bands at energies $E=\pm t$. Importantly, the dimerized SSH model
consists of two-site clusters that are completely disconnected from each
other. It is thus reasonable to expect the DC conductivity to
vanish.

As can be seen from Fig.~\ref{fig}, when the chemical
potential is tuned into the flat band, the interband conductivity
obtained from Eq.~\eqref{eq.kg_true} vanishes in the clean limit
$\eta\to 0^+$ in all cases. In the SSH model, the intraband
conductivity is exactly zero because the system contains only flat
bands, and the Kubo-Greenwood formula predicts a vanishing
$\sigma(\omega=0)$, Fig.~\ref{fig}b,c. In the sawtooth ladder (Fig.~\ref{fig}d,e), the intraband
contribution from the dispersive band is nonzero at any $T>0$, and
diverges in the limit $\eta\to 0^+$. However, it is highly suppressed
at low temperatures even for small $\eta$. At the values of
$\eta$ used here, the interband conductivity is dominant up to a
temperature $k_BT\sim 0.3$. This treshold temperature decreases with the
scattering rate
$\eta$. 

The conductivity obtained from Eq.~\eqref{eq.bastin}, the Kubo-Streda formula, is drastically
different. At exactly $T=0$, it retains a nonzero value proportional to the integrated quantum metric even when
$\eta\to 0^+$ (Fig.~\ref{fig}b,e). For $T>0$, the DC conductivity vanishes in the clean
limit in the dimerized SSH model (Fig.~\ref{fig}b). As a consequence, the limits $\lim_{T\to 0^+}$ and $\lim_{\eta\to
0^+}$ do not commute: taking $\lim_{T\to 0^+}$ before $\lim_{\eta\to
  0^+}$ gives a nonzero conductivity, whereas the inverted order
gives $\sigma(\omega=0)=0$. This $T=0$
behavior only exists when the chemical potential is tuned exactly into
the flat band: if it is slightly shifted away from the flat band, a
bump in $\sigma(\omega=0)$ appears at a nonzero temperature, as shown
in Fig.~\ref{fig}c, but the conductivity vanishes in the clean limit even
at $T=0$.

Even when the
clean limit given by Eqs.~\eqref{eq.bastin} and~\eqref{eq.kg_true} is the same, the behavior at finite scattering
rates is very different whenever a flat band is close to the Fermi
energy. In a dispersive band, however, both methods give close results
(see Fig.~\ref{fig}f). The discrepancy between the Kubo-Greenwood and
Kubo-Streda formulae is thus a flat-band effect. 

\subsection{Origin of the discrepancy}

In order to understand the differences in the results obtained from
Eq.~\eqref{eq.bastin} and Eq.~\eqref{eq.kg_true}, let us derive
Eq.~\eqref{eq.bastin} from Eq.~\eqref{eq.kg_true}. 

Starting from Eq.~\eqref{eq.kg_true}, we write
\eqa{
  \sigma_{\mu\nu}(\omega=0) &= -i\frac{e^2}{\hbar}\int_{\rm B.z.}\frac{d^D\vec{k}}{(2\pi)^D}\int_{-\infty}^{\infty}  
    d\epsilon \sum_{\vec{k}}\sum_{m\neq n} [j_{\mu}]_{nm}[j_{\nu}]_{mn} \nonumber\\
  &\left(
  \frac{n_F(\epsilon)\delta(\epsilon-\epsilon_n)}{(\epsilon-\epsilon_m)(\epsilon-\epsilon_m+i\eta)}
  - \frac{n_F(\epsilon)\delta(\epsilon-\epsilon_m)}{(\epsilon_n-\epsilon)(\epsilon_n-\epsilon+i\eta)}
  \right) , \label{eq.first_step}
}
Using that $\lim_{\eta\to 0^+}
1/[(\epsilon-\epsilon_n)(\epsilon-\epsilon_n+i\eta)] = -\lim_{\eta\to
  0^+}\partial_{\epsilon} (\epsilon-\epsilon_n+i\eta)^{-1}$, we obtain the Kubo-Bastin formula
\eqa{
  &\sigma_{\mu\nu}(\omega=0) = i\frac{e^2}{\hbar} \int_{\rm
    B.z.} \frac{d^D\vec{k}}{(2\pi)^D}
  \int_{-\infty}^{\infty} d\epsilon \:n_F(\epsilon) \nonumber\\ &{\rm Tr}
  \left[ j_{\mu} \frac{\partial G_{\vec{k}}(\epsilon+i\eta)}{\partial\epsilon}
    j_{\nu} \delta(\epsilon-H) - j_{\mu} \delta(\epsilon-H) j_{\nu}
    \frac{\partial G_{\vec{k}}(\epsilon-i\eta)}{\partial\epsilon}
    \right],
  \label{eq.kb_true}
} 
A detailed derivation of the full Kubo-Streda formula from this
form is given by Crépieux et al. in Ref.~\onlinecite{Crepieux2001}. Here, we will focus
on the symmetric part of the conductivity, which reads
\eqa{
  &\sigma_{\mu\nu}^s(\omega=0) = \frac{e^2}{\hbar} \int_{\rm B.z.}
  \frac{d^D\vec{k}}{(2\pi)^D} \int_{-\infty}^{\infty} 
    d\epsilon \: n_F(\epsilon)\times\nonumber\\ 
    &\times {\rm Tr} \big[
    j_{\mu}\frac{\partial}{\partial\epsilon} 
    {\rm Im}[G(\epsilon-i\eta)] j_{\nu} \delta(\epsilon-H) \nonumber\\
      &+
    j_{\mu} \delta(\epsilon-H) j_{\nu} \frac{\partial}{\partial\epsilon}
    {\rm  Im}[G(\epsilon-i\eta)]\big]
     \label{eq.bastin_step}
}
If we replace the delta functions by a Lorentzian, we can write
$\pi\delta(\epsilon-H) = \lim_{\eta'\to 0^+}{\rm Im} G(\epsilon-i\eta')$,
where $\eta'$ can generally be different from the scattering rate in
$G(\epsilon-i\eta)$. With $\eta'=\eta$, we obtain precisely the
formula~\eqref{eq.kg_wrong} through integration by parts. Taking $\eta'=\eta$ should not change the result in the clean limit as
long as the limit $\lim_{\eta'\to 0^+,\eta\to 0^+}$ does not depend on
the direction it is taken in. However, this is not always the case at $T=0$.
At exactly zero temperature, Eq.~\eqref{eq.bastin_step} becomes $\sigma_{\mu\nu}^s(\omega=0)=\sigma_{\mu\nu}^I+\sigma_{\mu\nu}^{II}$, where
\begin{widetext}
\eqa{
\sigma^I_{\mu\nu} &= \frac{e^2}{2\hbar\pi}\int_{\rm B.z.} \frac{d^D\vec{k}}{(2\pi)^D} 
{\rm Tr} \left[
    j_{\mu} {\rm
      Im}[G(\mu-i\eta)] j_{\nu} {\rm Im}[G(\mu-i\eta')] +
     j_{\mu} {\rm Im}[G(\mu-i\eta')] j_{\nu} {\rm
      Im}[G(\mu-i\eta)]
    \right] \\
    &= \frac{e^2}{2\hbar\pi}\sum_{mn}\int_{\rm B.z.} \frac{d^D\vec{k}}{(2\pi)^D} 
\left( \frac{\eta\eta'}{[(\epsilon_n-\mu)^2+\eta^2][(\epsilon_m-\mu)^2+\eta'^2]} + \frac{\eta\eta'}{[(\epsilon_n-\mu)^2+\eta'^2][(\epsilon_m-\mu)^2+\eta^2]} \right) [j_{\mu}]_{mn}[j_{\nu}]_{nm} \label{eq.integral}
    \\
\sigma^{II}_{\mu\nu} &= \frac{e^2}{2\hbar\pi}\int_{\rm B.z.} \frac{d^D\vec{k}}{(2\pi)^D}\int_{-\infty}^{\mu} d\epsilon \: {\rm Tr} \big[
    j_{\mu} \frac{\partial}{\partial\epsilon} {\rm
      Im}[G(\epsilon-i\eta)] j_{\nu} {\rm Im}[G(\epsilon-i\eta')] +
     j_{\mu} {\rm Im}[G(\epsilon-i\eta')] j_{\nu} \frac{\partial}{\partial\epsilon}{\rm
      Im}[G(\epsilon-i\eta)] \nonumber\\
    &\hphantom{= \frac{C}{2}\int_{\rm B.z.} \frac{d^D\vec{k}}{(2\pi)^D}\int_{-\infty}^{\mu} d\epsilon \: {\rm Tr} \big[}
      -
    j_{\mu} {\rm
      Im}[G(\epsilon-i\eta)] j_{\nu} \frac{\partial}{\partial\epsilon}{\rm Im}[G(\epsilon-i\eta')] -
     j_{\mu} \frac{\partial}{\partial\epsilon}{\rm Im}[G(\epsilon-i\eta')] j_{\nu} {\rm
      Im}[G(\epsilon-i\eta)]\big] .
}
\end{widetext}

The contribution $\sigma^I_{\mu\nu}$ causes the discrepancy between the Kubo-Greenwood and the Kubo-Streda formula in flat bands. Note that the double limit $\lim_{\eta\to 0^+,\eta'\to 0^+}$ of $\eta\eta'/[((\epsilon_n-\mu)^2+\eta^2)((\epsilon_m-\mu)^2+\eta'^2)]$ depends on the direction of the limits whenever only one of $\epsilon_n$ or $\epsilon_m$ is equal to $\mu$. For instance, if $\epsilon_n=\mu$ and $\epsilon_m\neq \mu$, the limit diverges if we first take $\eta\to 0^+$, vanishes if we take $\eta'\to 0^+$ before $\eta\to 0^+$, and gives the nonzero finite result $1/(\epsilon_m-\epsilon_n)^2$ if $\eta=\eta'$. The integrand in Eq.~\eqref{eq.integral} is thus problematic whenever one of $\epsilon_n$ or $\epsilon_m$ is equal to the Fermi energy. In other words, the contribution coming from states at the Fermi energy to the interband part of the conductivity can be inaccurate. 

In a dispersive band where the Fermi surface is $D-1$ dimensional, the states at exactly the Fermi energy will not contribute in the thermodynamic limit, since their area in the Brillouin zone vanishes. However, if the Fermi energy is in a (partially) flat band, these contributions appear in the final result. In particular, if we set $\eta=\eta'$ when taking the clean limit, the resulting DC conductivity when the chemical potential is tuned into a flat band is proportional to the integrated quantum metric of the flat band $n$,
\eq{
    \sigma_{\mu\nu} = \frac{e^2}{\hbar\pi}\int_{\rm B.z.} \frac{d^D\vec{k}}{(2\pi)^D} \sum_{m\neq n}
    \braket{\partial_{\mu}n_{\vec{k}}}{m_{\vec{k}}}\braket{m_{\vec{k}}}{\partial_{\nu}n_{\vec{k}}}.
}
However, this is only the case when taking the limit along $\eta=\eta'$, and the double limit is actually not well-defined. 

The source of the problem is the introduction of the derivative of the Fermi distribution. Above, we already mentioned that integrating $\partial n_F(\epsilon)/\partial \epsilon|_{\epsilon=\epsilon_n(\vec{k})}$ over the Brillouin zone causes problems when $\epsilon_n(\vec{k})$ is constant. Here, it may appear that the same problem is no longer present, because we could integrate $\partial n_F(\epsilon)/\partial\epsilon|_{\epsilon=\epsilon_n(\vec{k})}$ over the energy $\epsilon$. However, if we take $\eta'\to 0^+$ first before $\eta\to 0^+$, we recover the delta-function first introduced in Eq.~\eqref{eq.first_step}. Now that we already integrated over the energy to get rid of the delta function coming from $\partial n_F(\epsilon)/\partial\epsilon|_{\epsilon=\epsilon_n(\vec{k})}$, the remaining delta-function has transformed into $\delta(\mu-\epsilon_n(\vec{k}))$. Again, the integral over the Brillouin zone is no longer well-defined when the chemical potential is in a dispersionless flat band.

\section{Conclusions and discussion}\label{sec.conclusion}

We calculated the DC conductivity in a multiband system using the Kubo-Greenwood formula and the Kubo-Streda formula, and scrutinized various approximations used in the literature. Our focus was analyzing the DC conductivity in an isolated (gapped from other bands) flat band and its potential connection to quantum geometry. We summarize here our findings and discuss their implications. 

The Kubo-Greenwood formalism, without approximate use of derivatives of the Fermi function, predicts vanishing DC conductivity in a flat band in the clean limit $\eta \rightarrow 0$. This is physically intuitive considering that single particles have infinite effective mass and DC conductivity is essentially single particle transport in a system with no correlations but only a (vanishingly small) scattering rate $\eta$. The Kubo-Greenwood formula gives a vanishing DC conductivity in the clean limit in the dimerized SSH chain, consistent with the fact that transport through the chain is impossible since it is disconnected. In our view, the Kubo-Greenwood result of zero DC conductivity is the physically correct description of non-interacting electron transport in a flat band.

At finite $\eta$, the Kubo-Greenwood formula gives a DC conductivity related to the quantum metric, but this is simply the DC tail of the AC conductivity resonance at the band gap frequency. We argue that this contribution, proportional to $\eta$, is not physically meaningful in a flat band where $\eta$ cannot be claimed to be both finite and “small”, since there is no kinetic energy (band width) to compare it with. In other words, any small scattering, if considered to be truly finite, i.e.~$\eta \neq 0$, is actually a strong interaction/perturbation in a flat band. Thus, various approximations done in the derivation of the Kubo-Greenwood and related formulas, such as independent electrons, or neglecting vertex corrections, can be questioned.  

The Kubo-Streda formula gives results very different from the Kubo-Greenwood one in a flat band, and we argue they do not describe the limit of a dispersionless band correctly. This becomes apparent when they predict finite DC conductivity at zero temperature even in the completely disconnected SSH model where a DC current through the system clearly cannot flow. We point out in detail where the problems with this formula arise: essentially, they boil down to the lack of a Fermi surface. In a flat band, the Fermi energy is massively degenerate and forms a volume (3D) or a surface (2D) in momentum space. Therefore any unphysical features at the Fermi energy arising from approximations become finite, while in a dispersive band they would vanish within an integral over the Brillouin zone since there the Fermi surface is an area (3D) or a line (2D) of zero measure. One might think this could be solved by changing the integration variable from momentum to energy, but that would involve introducing the density of states which diverges in a flat band; the same problem dressed in a different way. Thus, in general, in studies of linear response phenomena in the flat band limit, one needs to be cautious with commonly applied approximations and formulas, since many of them are valid and physically meaningful only in the presence of a Fermi surface. Several recent studies of conductivity in a flat band~\cite{Mera2022,Bouzerar2022,Bouzerar2020,Bouzerar2021,Ferreira2015} should probably be revisited to understand the potential implications of our results there.

Our results relate also to the subtle connections between quantum geometry, physical observables, and the orbital positions in a lattice system. The symmetric and antisymmetric components of the quantum geometric tensor, namely the quantum metric and Berry curvature, depend on orbital positions (i.e.~change of orbital coordinates while keeping the connectivity (hopping) between the orbitals the same). It is well known that many physical observables, such as the anomalous Hall conductivity, are determined by the Berry curvature and thus depend on orbital positions. Flat band superconductivity was predicted to be proportional to the quantum metric (which is orbital dependent) within a large number of studies (see~\cite{Peotta2015,Torma2021} and references therein), while the definition of the superfluid weight is clearly independent of orbital positions. This discrepancy was solved only recently by showing that the superfluid weight is actually related to the minimal quantum metric, an orbital-independent quantity~\cite{Huhtinen2022}. Here we showed that the flat band AC conductivity can have a relation to the quantum metric, while the DC conductivity obviously is not related to any quantum geometric quantity as it vanishes. We believe that it is probably possible to determine, at a general level, which physical observables are orbital-position dependent simply based on Maxwell’s equations, general properties of the (superconducting or single electron) wavefunction, and gauge invariance. We leave this to a future work. Another important future research problem is to analyze transport in the case where other bands are touching the flat band.

\begin{acknowledgments}
We thank Johannes Mitscherling for useful discussions.
We acknowledge support by the Academy of Finland under the project number 349313.
K.-E.H. acknowledges financial support by the Magnus Ehrnrooth Foundation. 

\end{acknowledgments}

%


\end{document}